\documentstyle[preprint,eqsecnum,aps,psfig]{revtex}
\tightenlines
\def\beq{\begin{equation}}
\def\eeq{\end{equation}}
\def\bea{\begin{eqnarray}}
\def\eea{\end{eqnarray}}
\def\nnu{\nonumber}
\def\tst{\textstyle}

\def\fno#1{Fig.~\ref{#1}}

\def\eno#1{Eq.~(\ref{#1})}
\def\Slab#1{Sec.~\ref{#1}}

\def\al{\alpha}
\def\be{\beta}
\def\gam{\gamma}

\def\eps{\epsilon}

\def\kap{\kappa}

\def\sig{\sigma}

\def\Dta{\Delta}

\def\ptl{\partial}

\def\hf{{1\over2}}
\def\tshf{\tst\hf}
\def\quar{{1\over 4}}

\def\sixth{{1 \over 6}}

\def\lp{\left(}
\def\rp{\right)}

\def\ham{{\cal H}}
\def\ket#1{|#1\rangle}

\def\tran#1#2{\langle#1|#2\rangle}

\def\mel#1#2#3{\langle#1|#2|#3\rangle}

\def\bH{{\bf H}}
\def\bJ{{\bf J}}

\def\xhat{\bf{\hat x}}
\def\zhat{\bf{\hat z}}

\def\Fe8{Fe$_8$}
\def\hsc{{\ham_{\rm sc}}}

\def\dPh{\dot\Phi}
\def\ddPh{\ddot\Phi}

\def\cPh{\cos q}
\def\sPh{\sin q}
\def\ccPh{\cos 2 q}
\def\ssPh{\sin 2 q}
\def\dtc#1{{\dot t}_{#1c}}

\def\dy{\dot y}
\def\ddy{\ddot y}
\def\sss{\sig_1\sig_2}
\def\Ai{{\rm Ai}}
\def\Bi{{\rm Bi}}
\def\skc{\sinh\kap_c}
\def\ckc{\cosh\kap_c}

\begin{document}
\draft

\title{Discrete Phase Integral Method for Five-Term Recursion Relations}

\author{Anupam Garg$^*$}
\address{Department of Physics and Astronomy, Northwestern University,
Evanston, Illinois 60208}

\date{\today}

\maketitle

\begin{abstract}
A formalism is developed to study certain five-term recursion relations
by discrete phase integral (or Wentzel-Kramers-Brillouin) methods.
Such recursion relations arise naturally in the study of the
Schr\"odinger equation for certain spin Hamiltonians.
The conditions for the validity of the phase integral approximation are
derived. It is shown that in contrast to the three-term problem, it is now
possible to get a turning points ``under the barrier", i.e., in the
classically forbidden region, as well as inside the classically allowed
region. Further, no qualitatively new types of turning points arise in
recursion relations with still higher numbers of terms. The phase integral
approximation breaks down at the new turning points, requiring new
connection formulas, which are derived.
\end{abstract}

\widetext

\section{Introduction}
\label{intro}
The purpose of this paper is to develop the formalism of the discrete
phase integral (DPI), or Wentzel-Kramers-Brillouin method, for cases
where the recursion relation involves five terms. Previous use of
this method has, as far as we are aware, been limited to three-term
recursion relations \cite{dm,sg,pab,pab2,vs}. Surprisingly, the extension
to five terms is not routine, and entails novel physical and mathematical
considerations. Further, once this extension is understood, no
additional concepts are required in dealing with recursion relations
involving still more terms.

The physical problem which led the author to consider this extension
concerns the magnetic molecular cluster \cite{cs,alb,rc}
[(tacn)$_6$Fe$_8$O$_2$(OH)$_{12}$]$^{8+}$ (or just \Fe8 for short).
This molecule has a total spin $J=10$ in its ground state, and
crystallizes into a solid where the cluster has an approximate $D_2$
symmetry. The interaction between molecules is very weak, and
the low temperature spin dynamics of a single molecule are well
described by the Hamiltonian
\beq
\ham  = -k_2 J_z^2 + (k_1 - k_2) J_x^2 - g\mu_B \bJ\cdot\bH. \label{ham} 
\eeq
Here, $\bJ$ is a dimensionless spin operator, $\bH$ is an
externally applied magnetic field, and $k_1 > k_2 > 0$. (Experiments
reveal $k_1 \approx 0.33$ K, and $k_2 \approx 0.22$ K.)

The spectrum of the Hamiltonian (\ref{ham}) shows some extremely
interesting
features as a function of the applied field $\bH$. In particular, one
finds a large number of diabolical points \cite{bw} in the $H_x$-$H_z$
plane \cite{agepl,agdia,vf}, which have also been seen experimentally
\cite{ws}. Exactly as in the spectrum of a particle confined to a
triangular box \cite{bw}, some of the diabolical points (those arising
when $\bH \| \xhat$ or $\bH \| \zhat$) can be
related to a geometrical symmetry \cite{agprb}, but others can not.
The problem was first studied \cite{agepl} by instanton methods
when $\bH\|\xhat$, but this method is much harder to apply for
general field orientations, and the phase integral method proves
to be simpler. The existence of diabolical points turns out to depend
critically on having five terms in the recursion relation that we shall
describe shortly, and three terms would never lead to such points.
The calculations which pertain specifically to \Fe8 are described
elsewhere \cite{prl99}, but it seems worthwhile to present the
formal aspects of the work separately, as they are more generally
applicable.

\subsection{Heuristic discussion of the DPI approximation}
\label{heur}
It is useful to continue with the above example in order to introduce
the DPI method. The starting point of the procedure is to write
Schr\"odinger's equation
in the $J_z$ basis. Let $\ham\ket{\psi} = E \ket{\psi}$,
$J_z \ket m = m\ket m$, $\tran{m}{\psi} = C_m$,
and $\mel{m}{\ham}{m'} = t_{m,m'}$.  Then,
\beq
\sum_{n = m-2}^{m+2} t_{m,n} C_n = E C_m. \label{Seq}
\eeq
The diagonal terms ($t_{m,m}$) in the above equation arise
from the $J_z^2$ and $J_zH_z$ parts of $\ham$, those off-diagonal by
one ($t_{m,m \pm 1}$) from the $J_xH_x$ and $J_yH_y$ parts, and those
off-diagonal by two ($t_{m, m\pm 2}$) from the $J_x^2$ part.

The DPI method is applicable to a recursion relation such as
(\ref{Seq}) whenever the $t_{m,m\pm \al}$ ($\al = 0, 1, 2$)
vary sufficiently
slowly with $m$. A physical analogy may be made with an electron hopping
on a lattice with on-site energies $t_{m,m}$ and nearest-neighbor and
next-nearest-neighbor hopping terms $t_{m,m\pm 1}$ and $t_{m,m\pm 2}$.
If these quantities were independent of $m$, the solutions to \eno{Seq}
would be Bloch waves $C_m = \exp(iqm)$, with an energy
\beq
E = w_m + 2 t_{m,m+1} \cos q + 2 t_{m,m+2} \cos 2q
    \equiv E(q), \label{Evsq}
\eeq
where we have written $w_m \equiv t_{m,m}$ to highlight the physically
different role of the on-site energy from the hopping terms. We shall
use the notations $w_m$ and $t_{m,m}$ interchangably. If for fixed $\al$,
the $t_{m,m+\al}$ vary slowly with $m$ (where the meaning of this term
remains to be made precise), we expect it to be a good approximation to
introduce a local Bloch wavevector, $q(m)$, and write $C_m$ as an
exponential $e^{i\Phi}$, whose phase $\Phi$ accumulates approximately as
the integral of $q(m)$ with increasing $m$, in exactly the same way that
in the continuum quasiclassical method in one dimension, one writes
the wavefunction as $\exp(iS(x)/\hbar)$, and approximates $S(x)$ as the
integral of the local momentum $p(x)$.

It is obvious that the above approximation will entail the replacement
of various sums by integrals, and to that end, we introduce smooth
functions $t_{\al}(m)$ of a continuous variable $m$ as extensions of
$t_{m,m + \al}$ such that whenever $m$ is an integer \cite{fn1},
\beq
t_{\al}(m) = (t_{m,m+\al} + t_{m,m-\al})/2, \quad \al=0,1,2.
   \label{tcont}
\eeq
We will try and choose these functions so that their derivatives
are small. The precise way in which this is to be done will be discussed
later, but supposing that we have been successful in finding such
functions, we can seek to approximate $C_m$
in exact parallel with the continuum phase integral approach.
The form of the solution that emerges, and which readers will readily
appreciate from knowledge of the continuum case, is given by
\beq
C_m \sim {1 \over \sqrt{v(m)}}\exp\lp i\int^m q(m') dm'\rp,
    \label{Cwkb}
\eeq
where $q(m)$ and $v(m)$ obey the equations
\bea
E &=& w(m) + 2t_1(m) \cos q + 2t_2(m) \cos(2q)
     \equiv \hsc(q,m), \label{hjeq} \\
v(m) &=& \ptl \hsc/\ptl q = -2\sin q(m)
           \big(t_1(m) + 4 t_2(m) \cos q(m)\bigr).
     \label{vm}
\eea
[Just as for the matrix elements, we define $w(m) \equiv t_0(m)$,
and use the notation $w(m)$ when we want to emphasize it is as an
on-site energy.]
The interpretation of these equations is exactly the same as in the
continuum case. Thus, $\hsc(q,m)$ is a semiclassical Hamiltonian,
$q(m)$ is a local wavevector as already mentioned, and $v(m)$ is the
associated semiclassical electron velocity. We shall refer to 
Eq.~(\ref{Cwkb}) as the basic DPI form. Equation (\ref{hjeq}) is the
eikonal or Hamilton-Jacobi equation, while Eq.~(\ref{vm}) is the
discrete counterpart of the transport equation. The presence of the
lattice shows up in the $q$ dependence of $\hsc(q,m)$ through
periodic functions, whereas in the continuum case, such dependence is
typically of the form $q^2$.

As discussed by Braun \cite{pab,pab2}, the DPI approximation has been
employed in many problems in quantum mechanics where the Schr\"odinger
equation turns into a three-term recursion relation in a suitable
basis. All the types of problems as in the continuum case in
can then be treated---Bohr-Sommerfeld quantization, barrier
penetration, tunnel in symmetric double wells, etc. In addition, one
can also use the method to give asymptotic solutions for various
recursion relations of mathematical physics, such as those for the
Mathieu equation, Hermite polynomials, Bessel functions, and so on.
The general procedures are well known and simple to state. For any
$E$, one solves the Hamilton-Jacobi and transport equations to
obtain $q(m)$ and $v(m)$, and writes $C_m$ as a linear
combinations of the independent solutions that result. The interesting
features all arise from a single fact --- that the DPI approximation
breaks down at the so-called {\it turning points}. These are points
where $v(m)$ vanishes. One must relate the DPI solutions on opposite
sides of the turning point by connection formulas, and the solution
of all the various types of problems mentioned above depends on
judicious use of these formulas.

In this paper we will extend these ideas to five-term recursion
relations, focussing especially on those features which arise over
and above the three-term problem. Now, for any given $E$, there will
be {\it four} DPI solutions (\ref{Cwkb}), while in the three-term
case there are only two, because the Hamilton-Jacobi equation
(\ref{hjeq}) is a quartic in $e^{iq}$. These solutions will also
break down at turning points---points where $v(m)$ vanishes. In
contrast to the three-term case, we shall see that there are new types
of turning points. It is these turning points which are responsible for
the diabolical points in the spectrum of \Fe8. The three-term problem
turning points are analogous to those in the continuum quasiclassical
method, but the new ones that we will find are not. In fact, they can
only be described as lying ``under the barrier" from the continuum
viewpoint. These new or {\it irregular} turning points require new
connection formulas, which it is our goal to provide.

The plan of our paper is as follows. We will examine the 
DPI approximation carefully in \Slab{basic}, and see how it
fails when $v(m)$ vanishes. The precise width of the failure zone
is discussed in an Appendix. We will examine these failure or
turning points in \Slab{turns}, and see how the concept must be
extended beyond the three-term and continuum cases. We will find
that a turning point need not be a limit of the classically allowed
motion, and we will categorize the different types of turning points
that arise. We will conclude in \Slab{connect} by deriving
connection formulas at the new turning points.

We will limit ourselves to problems where the matrix $t_{m,n}$
is real and symmetric, $t_{m,n} = t_{n,m}$, as it simplifies the
analysis, and yet suffices to bring out all essential physical points.
In our spin example, this means that we only consider fields in the
$x$-$z$ plane. The extension to complex Hermitean matrices is cumbersome
to carry through, but presents no difficulty of principle. We shall
continue to couch our discussion in quantum mechanical language,
thinking of $E$ as an energy eigenvalue, although from the mathematical
viewpoint, this is not strictly necessary.

\section{The basic DPI approximation}
\label{basic}

In this section, we will examine the DPI approximation in more detail.
The argument proceeds in close analogy with the continuum case.
We begin by restating the approximation in a slightly different way
\cite{pab2,vf}.
Dividing \eno{Seq} by $C_m$, and writing $\zeta_{m+1} = C_{m+1}/C_m$,
we get
\beq
t_{m,m-2}\zeta_{m-1}^{-1} \zeta_m^{-1}
 + t_{m,m-1}\zeta_m^{-1}  + t_{m,m}
 + t_{m,m+1}\zeta_{m+1} + t_{m,m+2}\zeta_{m+1}\zeta_{m+2}
   = E. \label{ratio}
\eeq
If $t_{m,m+\al}$ for fixed $\al$ is almost the same over some range
$K \gg 1$
of $m$'s, then we will get almost the same numerical equation for the
$\zeta_m$'s over this range, and we will clearly obtain a good
approximate solution if we replace ratios like $\zeta_{m+1}/\zeta_m$
by unity. This leads to
\beq
t_{m,m-2}\zeta_m^{-2}
 + t_{m,m-1}\zeta_m^{-1}  + t_{m,m}
 + t_{m,m+1}\zeta_m + t_{m,m+2}\zeta_m^2
   = E, \label{ratio2}
\eeq
which is a solvable quartic equation in $\zeta_m$ (which is exactly
like the factor $e^{iq(m)}$ introduced in Sec.~\ref{intro}).
The corresponding approximation for $C_m$ is 
\beq
C_m \approx \prod_{k=m_a}^m \zeta_k, \label{prod}
\eeq
where $m_a$ is a suitable starting value. One could now use this
approximation to find the ratio $\zeta_{m+1}/\zeta_m$, substitute
this value for the ratio in \eno{ratio}, and solve again for $\zeta_m$.
The process could be further iterated if desired.

The product in \eno{prod} is more easily evaluated as a sum
by taking logarithms. Further, if the $\zeta_m$'s do not vary rapidly
from one $m$ to the next, the sum will be well approximated by an
integral. This makes it necessary to introduce continuum extensions
of the matrix elements $t_{m,m+\al}$. We turn therefore to this
problem, and discuss the condition for slow variation more clearly.
We are seeking functions $t_{\al}(m)$ such that \cite{fn1}
\beq
t_{\al}(m) = (t_{m,m+\al} + t_{m,m-\al})/2, \quad (\al=0,1,2),
   \label{tcont2}
\eeq
whenever $m$ is an integer. There are infinitely many such functions,
and we restrict them by imposing further conditions on a certain
number of their higher derivatives. For $t_1(m)$, e.g., we could
also demand (using dots to denote derivatives with respect to $m$),
\bea
{\dot t}_1(m) &=& t_{m,m+1} - t_{m,m-1}, \label{dott1} \\
{\ddot t}_1(m) &=& \tshf\left(
          t_{m+1,m+2} - t_{m,m+1} - t_{m,m-1} + t_{m+1,m+2}\right).
                \label{ddott1}
\eea
Similar conditions can be imposed on $t_0(m)$ and $t_1(m)$.
In general, we need conditions only up to some small order for
practical applications. Up to the degree of approximation in
\eno{Cwkb}, e.g., we need only go up to second derivatives.

Since the matrix elements are assumed to vary slowly, we want these
derivatives to be small, and this condition is best codified in terms
of a small parameter $\eps$, which plays the same role as $\hbar$ in
the continuum case, such that $t_{\al}$ is formally of order
$\eps^0$, ${\dot t}_{\al}$ of order $\eps$, ${\ddot t}_{\al}$ of
order $\eps^2$, and so on. For spin Hamiltonians such as \eno{ham},
this parameter is $1/J$. The quasiclassicality conditions then read
\beq
{dt_\al \over dm} = O\left( {t_{\al}(m) \over J } \rp, \quad
{d^2t_\al \over dm^2} = O\left( {t_{\al}(m) \over J^2 } \rp,
\label{slow}
\eeq
etc. We will continue to use $1/J$ as a generic small parameter 
in the rest of our analysis. A problem in which one can not find
functions $t_{\al}(m)$ obeying \eno{slow} will not be amenable
to a phase integral approximation.

One small point should be kept in mind while judging orders of
smallness in the spin problem, and others like it. There are natural
algebraic expressions for the $t_{\al}(m)$'s in which $m$ appears only
in the combination $m/J$ or $m/[J(J+1)]^{1/2}$. Thus, derivatives
with respect to $m$ are automatically smaller by an order $J^{-1}$.
As $J\to\infty$, however, the classical quantity is not $m$, but
$m/J$. Thus the ratio $m/J$ should be regarded as a quantity of
order unity and not $J^{-1}$.
 
With this lengthy preamble, we are ready to solve our recursion
relation. As in the continuum case, we make the exponential
substitution,
\beq
C_m = e^{i\Phi(m)}. \label{eiPhi}
\eeq
The first approximation is obtained if we 
assume that $\dPh(m)$ varies slowly so that $\ddPh(m)$ may be neglected.
Then $C_{m\pm \al} \approx C_m e^{i\al \dPh}$, and substituting this into
the recurrence relation (\ref{Seq}) with
$t_{m,m\pm\al} \approx t_{\al}(m)$, we obtain the Hamilton-Jacobi
equation (\ref{hjeq}) with $\dPh(m) = q(m)$. 

To proceed more systematically, we look for a solution for
$\Phi(m)$ as a series in inverse powers of $J$:
\beq
\Phi = \Phi_0 + \Phi_1 + \Phi_2 + \cdots, \label{Phi012}
\eeq
where, 
\bea
\Phi_n &=& O(J^{1-n}), \nnu \\
\dot\Phi_n &=& O(\Phi_n/J), \label{Phidot} \\
\ddot\Phi_n &=& O(\Phi_n/J^2), \nnu
\eea
and so on. The successive inverse powers of $J$ in the derivatives are expected since
we expect the $\Phi_n$ to be simple functions of the $t_{m,m+\al}$, and we shall soon
see whether or not this expectation is fulfilled. $\Phi_0$ is our zeroth order
approximation above, and so we set
\beq
\dPh_0(m) = q(m) \label{tP0}
\eeq
from the outset. As in the continuum case, we need to keep terms up to
$\Phi_2$ in order to decide if the approximation is succeeding \cite{Bohm}.

Up to terms of order $1/J^2$ relative to the leading one, we have
\bea
C_{m\pm 1} 
  &=& C_m \exp \left[i\lp \pm q \pm \dPh_1 \pm \dPh_2
                       + \hf( \dot q + \ddPh_1)
                         \pm \sixth \ddot q + \cdots \rp\right] \nnu \\
  &=& C_m e^{\pm i q} \left[
             1 + {i \over 2} (\dot q \pm 2 \dPh_1)
               - {1\over 8} (\dot q \pm 2 \dPh_1)^2
               \pm i\dPh_2 + {i\over 2} \ddPh_1
               \pm {i \over 6} \ddot q + \cdots \right].
\eea
We now wish to substitute this form into our recursion relation.
We would like to use the continuum forms $t_{\al}(m)$ instead of
the discrete matrix elements $t_{m,m+\al}$. Just as there are
infinitely many continuous functions which we could take, there 
are many approximants for $t_{m,m+\al}$. Note in particular, that
there is no unique way to ``solve" Eqs.~(\ref{tcont2})--(\ref{ddott1})
for the matrix elements in terms of the continuous functions.
The simplest procedure is to take $t_{m,m} = t_0(m)$,
$t_{m,m\pm 1} = t_1(m \pm \tshf)$, and
$t_{m,m\pm 2} = t_2(m \pm 1)$. For $t_{m,m\pm 1}$, in particular,
a Taylor expansion of this approximation gives
\beq
t_{m,m\pm 1} = t_1(m) \pm \hf {\dot t}_1(m)
               + {1 \over 8} {\ddot t}_1(m) + \cdots, \label{tder}
\eeq
where the error is of order $J^{-3}$. Therefore,
\beq
\sum_{n = m\pm 1} t_{m,n} C_n
   =  t_1(m) \al_1(m) + {\dot t}_1(m) \al_2 (m)
         + {\ddot t}_1(m) \al_3(m) + \cdots, \label{t1term}
\eeq
where
\bea
\al_1(m) &=& 2 C_m \biggl[ \cPh
               + {i \over 2} \lp \dot q \cPh + 2i \dPh_1 \sPh \rp
   -{1 \over 8} \lp {\dot q}^2 + 4 \dPh_1^2 -4i \ddPh_1 \rp \cPh\biggr. \nnu \\
   &\quad& \biggl. -{i \over 6} \lp 3 \dot q \dPh_1
             - 6i \dPh_2 - i \ddot q \rp\sPh \biggr], \nnu \\
\al_2(m) &=& i C_m \left[ \sPh +{i \over 2} \dot q \sPh
                +  \dPh_1\cPh \right],
              \label{alphas} \\
\al_3(m) &=& \quar C_m \cPh. \nnu
\eea
Note that we have kept only terms up to order $1/J^2$ in $\al_1$,
$1/J$ in $\al_2$, and $J^0$ in $\al_3$, since relative to $t_1(m)$,
${\dot t}_1$ and ${\ddot t}_1$ are of order $1/J$ and $1/J^2$,
respectively.

Similarly,
\beq
\sum_{n = m\pm 2} t_{m,n} C_n
   =  t_2(m) \be_1(m) + {\dot t}_2(m) \be_2 (m)
         + {\ddot t}_2(m) \be_3(m) + \cdots, \label{t2term}
\eeq
where
\bea
\be_1(m) &=& 2 C_m \biggl[ \ccPh
               + 2i \lp \dot q \ccPh + i \dPh_1 \ssPh \rp
   - 2 \lp {\dot q}^2 +  \dPh_1^2 - i \ddPh_1 \rp \ccPh \biggr.\nnu \\
   &\quad& \biggl.-{2i \over 3}
      \lp 6 \dot q \dPh_1 - 3i \dPh_2 - 2i \ddot q \rp\ssPh \biggr], \nnu \\
\be_2(m) &=& 2i C_m \left[ \ssPh + 2i \dot q \ssPh + 2 \dPh_1\ccPh \right],
              \label{betas} \\
\be_3(m) &=& C_m \ccPh. \nnu
\eea

We now substitute these relations into the recurrence relation
(\ref{Seq}), and equate equal powers of $J$. The terms of order $J^0$
obviously give Eq.~(\ref{hjeq}), while those of order $J^{-1}$ give,
after some work,
\beq
-2\dPh_1 (t_1\sin q + 2t_2 \sin 2q) 
          = -i {d \over dm}\lp t_1\sin q + 2t_2 \sin 2q \rp.
  \label{dP1}
\eeq
The left hand side of this equation is $\dPh_1 v(m)$ [see
\eno{vm}], while the right hand side may be written as
$(i/2)d v(m)/dm$. Thus, integration yields (after a suitable
choice of an indefinite constant)
\beq
\Phi_1(m) = {i \over 2} \ln v(m). \label{defP1}
\eeq
Note that as assumed in \eno{Phidot}, $\Phi_1 = O(J^0)$.
Pending a demonstration that $\Phi_2(m)$ is negligible, we have arrived
at the basic DPI form (\ref{Cwkb}), which we now see as the first two terms
of an asymptotic expansion of $\Phi$ in inverse powers of $J$.

The equation for $\Phi_2(m)$ is considerably more involved. After
some analysis, we find
\beq
{d {\tilde\Phi}_2 \over dm} =
   {1\over 8} {d^2 r \over dm^2}
        + {i \over 4} r {d \dPh_1 \over dm}, \label{dP2}
\eeq
where we have defined
\bea
&&{\tilde\Phi}_2 = \Phi_2 +
     {1\over 24}\int {t_1 + 16t_2 \cos q \over t_1 + 4t_2 \cos q}
       \ddot q\, dm, \label{P2til} \\
&& r(m,q(m)) = {t_1 \cos q + 4t_2 \cos 2q
                  \over t_1 \sin q + 2t_2 \sin 2q}. \label{defr}
\eea
Thus,
\beq
\Phi_2 = 
     - {1\over 24}\int {t_1 + 16t_2 \cos q \over t_1 + 4t_2 \cos q}
       \ddot q\, dm
     + {1 \over 8} {dr \over dm}
     - {1 \over 8} \int r {d^2 \over dm^2} \ln v \, dm.
            \label{ansP2}
\eeq
Based on power counting, this is indeed of order $J^{-1}$, since
${\ddot q}$ and $d^2 \ln v(m)/dm^2$ are of order $J^{-2}$. Thus,
$|\Phi_2| \ll \Phi_1 = O(J^0)$. However, it is plain that this
condition is violated whenever $v(m)$ approaches zero, for then
both $\ln v(m)$ and $r(m)$ diverge. To find the actual magnitude
of $\Phi_2$, we need to know how $q(m)$ and $v(m)$ behave near a
turning point. This behavior is found in the next section [see
Eqs.~(\ref{qntp}) and (\ref{vntp})]. The magnitude of $\Phi_2$
is estimated in Appendix \ref{valid}, where we show that the width
of the zone where DPI fails is of order $J^{1/3}$.
 
\section{Turning points}
\label{turns}
We now turn to a study of the points where
$v(m) = \ptl\hsc(q,m)/\ptl q = 0$.  We shall call all such points
{\it turning points} in analogy with the continuum case. In contrast
to that case, however, we will find that turning points
are not just the limits of the classical motion for a given energy,
once the notion of the classically accessible region is suitably
understood.

Since we must also obey the eikonal equation (\ref{hjeq}) in addition
to the condition $v(m) = 0$, at a turning point both $m$ and $q$ are
determined if $E$ is given.  Setting $v=0$ in \eno{vm}, we see that we
must have either $q=0$, or $q=\pi$, or $q = q_*(m)$, where
\beq
\cos q_*(m) = - t_1(m)/4t_2(m). \label{cosqstar}
\eeq
Substituting these value of $q$ in the eikonal equation, we see that
a turning point arises whenever
\beq
E = U_0(m),\ U_{\pi}(m),\ {\rm or}\ U_*(m),
   \label{Econd}
\eeq
where,
\bea
U_0(m) &=& \hsc(0,m) = w(m) + 2t_1(m) + 2t_2(m), \label{U0} \\
U_{\pi}(m) &=& \hsc(\pi, m) = w(m) - 2t_1(m) + 2t_2(m), \label{Upi} \\
U_*(m) &=& \hsc(q_*,m) = w(m) - 2t_2(m) - {t_1^2(m) \over 4 t_2(m)}.
           \label{Ustar}
\eea
Note that $q_*(m)$ may be complex for some $m$, but since
$\cos q_*$ is always real, $U_*$ is real for all $m$. We shall
refer to these three energy curves as {\it critical curves}.

In the one-dimensional continuum case where $\hsc = (p^2/2m) + V(x)$,
the condition $v(x) = 0$ is equivalent to $E = V(x)$. The latter
condition marks the edge of the classically allowed region,
$E < V(x)$. Let us recall why this is said to be so. A particle at a
point $x$ where $V(x) > E$ must be ascribed an imaginary momentum. To
understand the analogous condition in the discrete case, let us return
to the analogy of an electron in a one-dimensional
lattice with $m$-independent matrix elements $t_{m,m+\al}$. Equation
(\ref{Evsq}) gives the dispersion relation for an energy band $E(q)$.
The classicaly allowed range of energies is now defined by the limits
of this band since (provided one is not too close to a band edge) a
spatially localized electron with a mean energy in this range can be
constructed as a wavepacket out of Bloch states with only real
wavevectors. Alternatively, we could say that solutions with energy in
the allowed range correspond to travelling waves, while solutions
outside this range correspond to evanescent waves. These notions
continue to be valid when $t_{\al}(m)$ are slowly varying with
$m$. The dispersion relation may be taken as $\hsc(q,m)$ for
any fixed value of $m$. The band is now $m$-dependent, and so in
particular are the band edges, which we denote by $U_-(m)$ (lower
edge) and $U_+(m)$ (upper edge).

To find the band edges, we note that by definition, the energy at
an edge is either a minimum or maximum, so $\ptl\hsc(q,m)/\ptl q = 0$,
i.e., $v(m) =0$. [The converse is not true, i.e., $v(m)= 0$ need not
always define a band edge.] Thus the wavevectors at the band edges 
are again to be found in the set $q = 0$, $\pi$, and $q_*$, with the
answers depending on the signs and magnitudes of $t_1$ and $t_2$.
[We do not distinguish between $q=\pi$ and $q= - \pi$, or between
$q=q_*$ and $q=-q_*$, as $\hsc(q,m)$ is an even function of $q$.]
To narrow down the number of cases to be considered, we observe that
the gauge transformation $C_m \to (-1)^m C_m$ changes the sign of
$t_1$. Hence, we may assume $t_1 < 0$ without loss of generality.
Consider the case $t_2 > 0$ first. Then the upper band edge
is always located at $q=\pi$, while the lower band edge
is located at $q=0$ if $t_1/4t_2 < -1$, and at
$q_* = \cos^{-1}(-t_1/4t_2)$ if $-1 < t_1/4t_2< 0$. The case $t_2 < 0$
is completely analogous. Now the lower band edge is always at $q=0$,
while the upper band edge is at $q=\pi$ for $t_1/4t_2 > 1$, and at
$q_* = \cos^{-1}(-t_1/4t_2)$ for $ 0 < t_1/4t_2 < 1$. The
different types of band energy curves that can arise are illustrated
in \fno{bands}. In the rest of this paper we shall carry out the
analysis in detail assuming $t_2 >0$. The case $t_2 < 0$ is easily
treated in parallel, and we shall only give the final results where
these are significantly different.

The conditions \eno{Econd} for a turning point
are analogous to the requirement that
$E = V(x)$ in the continuum case. Correspondingly, it helps in
visualization to draw all three critical curves versus $m$, and a
horizontal line indicating the energy. Any intersection of this line
with a critical curve is a turning point. See, e.g.,
Figs.~\ref{ccurves} and \ref{critfe8}.

Not every critical curve need be a band edge curve, however.
Since we have chosen $t_1(m) < 0$ and $t_2(m) > 0$,
the upper edge curve $U_+(m)$ is the same as $U_{\pi}(m)$ for all $m$,
but the lower edge curve, $U_-(m)$, may be $U_0(m)$ for some values of $m$,
and $U_*(m)$ for other values as discussed above. It turns out to be
useful to introduce a dual labelling scheme for the critical curves
and write
\bea
U_0(m) = U_i(m),\  U_*(m) = U_-(m),\quad{\rm if}\ q_* \in (0,\pi),
                                             \label{Ui} \\
U_0(m) = U_-(m),\  U_*(m) = U_f(m),\quad{\rm if}\ q_* \not\in (0,\pi).
                                             \label{Uf} 
\eea
We have already noted that $U_{\pi}(m) = U_+(m)$. 
The subscripts $i$ and $f$ stand for ``internal" and ``forbidden", since
in the first case above, $U_0(m)$ lies inside the classically allowed
energy range, while in the second case, $U_*(m)$ lies outside this range.
The turning points with $E = U_i$ and $E = U_f$ have no analogues in
continuum quantum mechanical problems.

Before turning to a classification of the various turning points,
however, it is useful to record some further properties of the critical
curves. The first property is that $U_0(m) \ge U_*(m)$, since
\beq
U_0(m) - U_*(m) = {1 \over 4t_2(m)}
        \bigl(t_1(m) + 4t_2(m) \bigr)^2. \label{difU}
\eeq
Differentiating this equation with respect to $m$, it follows that the
case of equality, $U_0(m) = U_*(m)$, happens at a point where both curves
have a common tangent. These facts are illustrated in \fno{fe8critmag}.
Further, at the point of contact, which we denote by $m^*$,
$t_1(m)/4t_2(m) = -1$, which is precisely the condition derived above
for the lower band edge to change from $q=0$ to $q=q_*$.

The second property provides an alternative way of viewing the
condition $E = U_*(m)$. Solving the eikonal equation (\ref{hjeq}) for
$\cos q$ we obtain
\bea
\cos q(m) &=& {-t_1(m) \pm [t^2_1(m) - 4t_2(m) f(m)]^{1/2}
              \over 4t_2(m)}; \label{cosq} \\
f(m) &=& w(m) - 2t_2(m) -E. \label{fofm}
\eea
Since $\cos q = -t_1/4t_2$ at $q=q_*$, the discriminant in \eno{cosq}
must vanish, and we must have
\beq
t_1^2(m) = 4t_2(m) \bigl( w(m) - 2t_2(m) - E \bigr)
           \qquad (q = q_*). \label{tp2}
\eeq
It is easily verified that this equality is identical to
$E = U_*(m)$.

We now turn to discussing the different types of turning points:

{\it Type} $A$: $E = U_-(m)$ when $U_- = U_0$. See, e.g.,
\fno{ccurves}. The
region $m \le m_c$ is classically allowed.  This is analogous to what
happens in the conventional continuum quasiclassical method --- the
turning point is located at the boundary of the classically accessible
region for the energy given. For $m$ just less than $m_c$, there are
two solutions of the Hamilton-Jacobi equation (\ref{hjeq}) with
$q \approx [\bigl(E - U_-(m)\bigr)/a]^{1/2}$, where
$a = -(t_1 + 4t_2)$. For $m$ just greater than $m_c$, these values
of $q$ continue on to the imaginary axis. The corresponding
wavefunctions $C_m$ change from slowly oscillatory for $m<m_c$ to
exponentially growing and decaying for $m > m_c$. The connection formulas
for these solutions are exactly like those in the continuum case,
and may be derived as in Refs.~\cite{sg,pab}.
Note that the other two solutions of the
Hamilton-Jacobi equation evolve smoothly, and the corresponding DPI
wavefunctions $C_m$ do not need to be ``connected" across this turning
point.

{\it Type} $\bar A$: $E = U_+(m)$. See \fno{ccurves} again.
The region $m \ge m_b$ is classically allowed. This case is physically
very similar to type $A$ in that the turning point is at the boundary
of the classically allowed and forbidden regions. Now, however,
$q \approx \pi$ in the transition zone, so the wavefunctions 
contain a rapidly oscillating factor $(-1)^m$ in addition to all the
other variation. Although the connection formulas can be derived from
those for type $A$ turning points by means of the transformation
$C_m \to (-1)^m C_m$, as shown in Ref.~\cite{pab}, their detailed form
has a very different superficial look.

{\it Type} $A'$: $E = U_i(m)$. Consider \fno{critfe8}, and
the energy $E$ shown there, which intersects $U_i(m)$ at
$m=m_c$. In \fno{Aprband} we sketch energy bands for this problem for
several
values of $m$. For $m$ just less than $m_c$, $\hsc(q,m) = E$ in just
two places, which we denote by $\pm q(m)$. (We do not show the solution
$-q(m)$ explicitly.) For $m$ just greater than $m_c$, two new
intersections develop at $\pm q'(m) \approx 0$. Thus for $m>m_c$, our
wavefunction consists of a sum of four basic solutions (\ref{Cwkb}),
all oscillatory, while for $m<m_c$ we have two oscillatory solutions
[associated with $\pm q(m)$], and two exponentialy decaying or growing
solutions [associated with $\pm q'(m)$]. The latter solutions must be
related across the turning point by connection formulas, which are
completely identical to those for type $A$. The fact that we may
have exponentially decaying and growing solutions inside a classically
allowed region is unexpected from prior experience with the continuum
quantum mechanical problems, and underscores the point that this
turning point has no analogue there. We are unaware if it has ever been
considered in other physical situations where a continuum phase
integral approach may be applied.

{\it Type} $B$: $E= U_f(m)$. This turning point is perhaps the most
interesting of all. Since the energy now lies outside the classically
allowed range, the point lies ``under the barrier", and at it, $q$ must
be purely imaginary. Consider \fno{fe8critmag}, and the energy $E$,
which intersects $U_f$ at $m = m_c$. We see that for $m \le m_c$,
there are two solutions to the Hamilton-Jacobi equation (\ref{hjeq}) with
$q = \pm i\kap$, where $\kap$ is real. For $m > m_c$, these solutions
acquire a real part as well, so that $C_m$ [See Eq.~(\ref{Cwkb})] changes
from a decaying (or growing) exponential to an oscillating solution with
an exponentially decaying (or growing) envelope. As for
type $A'$ points, this behavior has no analogue in continuum quantum
mechanical problems, and we are unaware of prior analyses in other
contexts. We shall derive connection formulas for this case in
Sec.~\ref{connect}.

{\it Type} $B'$: $E = U_-(m)$ when $U_-(m) = U_*(m)$. This turning
point is like type $A$ in that the energy lies at the lower limit of the
classically allowed region, but like type $B$ in that $q \ne 0$.
The solutions which must be connected are purely oscillatory on the
classically allowed side, with $q \simeq \pm q_*$, and oscillatory
exponentials (growing or decaying) on the forbidden side. The
connection formulas are similar to those for case $B$.

Our nomenclature for the turning points may have become evident to
the reader. The letter $A$ indicates that $q =0$ or $\pi$, while $B$
indicates that $q = q^*$. A bar designates cases where $q$ has a value
close to $\pi$, leading to a oscillatory factor in $C_m$ close to
$(-1)^m$, and a prime indicates cases where $U_*$ is either $U_-$
or $U_+$ at the turning point value of $m$. Thus in the
case $t_1 < 0$, $t_2 < 0$, we would have turning points of type $A$,
$\bar A$, ${\bar A}'$, $\bar B$, and ${\bar B}'$. Note that the
mathematical aspects of the turning point, i.e., connection formulas,
are governed by the value of $q$ at the turning point, but its physical
nature is governed by whether the energy lies at the boundary (cases
$A$, $\bar A$, $B'$, and ${\bar B}'$), in the interior (cases $A'$
and ${\bar A}'$), or in the exterior (cases $B$ and $\bar B$) of the
classically allowed range.

It will now be apparent that in problems with three-term recursion
relations, where $\hsc(q,m) = w(m) + 2t_1(m)\cos q$, and
$v(m) = -2t_1(m)\sin q(m)$, the only turning points are at band edges,
with $q = 0$ or $\pi$, i.e., of type $A$ or $A'$. It is also apparent
that no new points are involved in problems with further neighbor
hopping, i.e., recursion relations with seven or more terms. Turning
points are encountered whenever the energy lies on a critical curve,
where this term now describes all curves in the $E$-$m$ plane on which
$\ptl \hsc(q,m)/\ptl q = 0$. As $m$ is varied through each turning
point, two roots of the Hamilton-Jacobi equation for $q$ approach
each other parallel to either the real or imaginary axis, coalesce,
and move apart in the orthogonal direction \cite{fn3}. Consider for
example the band structure in \fno{strband}, which could arise from a
recursion relation with seven or more terms. The accompanying critical
curves are shown in \fno{strcrit}. For the energy $E$ shown, if
$m < m_a$, there is only one real solution for $q$, i.e. $q_+$. (We do
not explicitly mention negative values of $q$.) At $m=m_a$, a new real
value of $q$, $q_2$, enters the picture, and splits into two solutions
$q_{2-}$ and $q_{2+}$ as $m$ increases further.

We conclude this section by finding the general behavior of $q(m)$
and $v(m)$ near a turning point $m=m_c$. In its vicinity we may write
\beq
\hsc(q,m) = \hsc(q,m_c) +
    \left.(m-m_c){\ptl \hsc \over \ptl m}\right|_{m_c}
        + \cdots.
     \label{htaylor}
\eeq
By definition, however, $v(m_c) = 0$, and so $\hsc(q,m) - E$,
and $\ptl[\hsc(q,m_c)-E]/\ptl q$ {\it both} vanish. If we write
$q(m_c) = q_c$, expand the the right hand side of Eq.~(\ref{htaylor})
in powers of $q-q_c$
as well, and retain only the leading non vanishing terms in
$q-q_c$ and $m-m_c$, we obtain
\beq
\hsc(q,m) - E \approx a(q-q_c)^2 + b(m-m_c) + \cdots,
   \label{hqm}
\eeq
where $a$ and $b$ are constants. If we regard $w(m)$ and $t_{\al}(m)$
as being of order $J^0$, then by Eq.~(\ref{slow}), we have $a=O(J^0)$,
and $b = O(1/J)$, and therefore
\bea
q(m) - q_c &\sim& [(m-m_c)/J]^{1/2}, \label{qntp} \\
v(m)       &\sim& [(m-m_c)/J]^{1/2}. \label{vntp}
\eea
These formulas prove useful when connection formulas are derived.
Their usefulness is limited, however, if there is another turning
point very close to $m_c$. In this case we should keep terms of
order $(m-m_c)^2$. The requisite analysis is very similar to that of
quadratic turning points \cite{bm} in the continuum case, but we
shall not have any occasion to pursue it further. In all the
calculations we have done for the \Fe8 or other spin problems
\cite{agjmp}, we have been able to sidestep the associated
quadratic connection formulas by directly matching the solutions
in the forbidden region to solutions of the Schr\"odinger equation
for a harmonic oscillator.

\section{Connection Formulas for Forbidden Region Turning Points}
\label{connect}

We turn at last to the problem of finding connection formulas at
the turning points. The formulas for points of type $A$, $\bar A$, etc.
are quoted by Braun \cite{pab}, so we will only consider points of
type $B$ and $\bar B$. Our procedure is a small modification of that
used by Schulten and Gordon \cite{sg}. Suppose the turning point is
at $m=m_c$, and $q(m_c) = q_c$. As shown in Appendix \ref{valid},
the DPI solution (\ref{Cwkb}) fails in a window
$\Dta m  \equiv |m - m_c| \le O(J^{1/3})$, which we shall refer to as
the failure zone. The first
step is therefore to find another approximation that holds in the
larger window (which we refer to as the central zone)
$\Dta m  \ll J^{\eta}$, where $\eta > 1/3$. To do
this we write $C_m$ as an $e^{iq_c m}$ times a slowly varying factor,
$y_m$, for which we then derive an approximate second order differential
equation. The second step is to asymptotically match solutions of this
differential equation to the DPI solutions in the overlap zones
$J^{\eta} \gg \Dta m \gg J^{1/3}$ on
either side of the turning point where both types of solutions are valid.
The last step is to directly write down the transformation
matrix between the coefficients of the linear combination of DPI solutions 
for $m<m_c$ to those for $m>m_c$, without having to consider the
solution in the intermediate zone. We will carry out these three
steps only to an order necessary to match the solutions to the
accuracy represented by Eq.~(\ref{Cwkb}), i.e., to order $J^0$ in the
phase $\Phi(m)$ introduced after Eq.~(\ref{Seq}) or in Eq.~(\ref{eiPhi}).

Let us assume as before that $t_1 <0$, $t_2 > 0$, and first consider
turning points of type $B$. We will denote quantities evaluated at
$m=m_c$ by a subscript $c$: $t_1(m_c) = t_{1c}$, ${\dot t}_1(m_c)
= \dtc1$, etc. Let $q$ be pure imaginary for $m \le m_c$, and
precisely at $m_c$ let us write
\beq
q(m_c) = i\sig_2 \kap_c, \label{defkc}
\eeq
where $\kap_c  >0$ and $\sig_2 = \pm 1$. Putting this in
Eqs.~(\ref{cosqstar}) and (\ref{hjeq}) we have
\bea
t_{1c} &=& - 4 t_{2c}\cosh\kap_c. \label{t1c} \\
E &=& w_c + 2 t_{1c} \cosh\kap_c + 2 t_{2c} \cosh 2\kap_c. \label{Eatmc}
\eea

To carry out step 1, we write
\beq
C_m \approx e^{-\sig_2 \kap_c (m - m_c)} y(m), \label{Ctoy}
\eeq
where ${\dot y}(m) \ll \kap_c$. Assuming that this is so, we write
\beq
C_{m \pm k} = e^{-\sig_2 \kap_c (m - m_c)} e^{-\sig_2 k \kap_c}
                [y  \pm k\dy + \hf k^2 \ddy + \cdots]. \label{Cpmk} 
\eeq
This approximation will hold provided $\dot y \ll y$, and
$\ddot y \ll \dot y$ throughout the central zone
$|m - m_c| \ll J^{\eta}$. [We anticipate, in fact,
that throughout this zone, 
${\dot y}(m) \sim J^{-\gam}y(m)$ where
$\gam > 0$.] That this is so and that 
higher order derivatives can be neglected in Eq.~(\ref{Cpmk})
will be justified
post facto.  Substituting Eq.~(\ref{Cpmk}) in
Eq.~(\ref{Seq}) we obtain
\beq
A_0(m) {\ddy}(m) + A_1(m){\dy}(m) + A_2(m)y(m) \approx 0, 
    \label{diffy}
\eeq
where
\bea
A_0 &=& \sum_{k=\pm1, \pm2} \hf k^2 e^{-\sig_2 k\kap_c} t_{m,m+k}.
                   \label{A0} \\
A_1 &=& \sum_{k=\pm1, \pm2} k e^{-\sig_2 k\kap_c} t_{m,m+k}, 
                   \label{A1} \\
A_2 &=& \sum_{k=\pm1, \pm2} e^{-\sig_2 k\kap_c} t_{m,m+k}
        + w_m - E, \label{A2}
\eea

The differential equation that we are seeking for $y(m)$ need only hold
in window of width
$O(J^{\eta})$ with $\eta > 1/3$ around $m_c$. If we choose $\eta <1$,
we can use the fact that $w(m)$ and $t_\al(m)$ are slowly varying,
and use the expansions
\bea
t_{m,m\pm k} \approx t_k(m) \pm \hf k\, {\dot t}_k(m), \label{tmm} \\
t_k(m) \approx t_{kc} + (m-m_c)\dtc{k}, \label{ttdot}
\eea
etc. Doing this, we obtain
\bea
A_0 &\approx& a_1, \label{A0a1} \\
A_1 &\approx& a_2 -\sig_2 (m - m_c) b_2, \label{A1a2} \\
A_2 &\approx& -\sig_2 a_3 + (m - m_c) b_3, \label{A2a3}
\eea
where
\bea
a_1 &=& t_{1c}\cosh\kap_c + 4 t_{2c} \cosh 2\kap_c, \label{a1} \\
a_2 &=& \dtc1\cosh\kap_c + 4 \dtc2 \cosh 2\kap_c, \label{a2} \\ 
b_2 &=& 2 \sinh\kap_c ( \dtc1 + 4\dtc2 \cosh\kap_c), \label{b2} \\
a_3 &=&  \sinh\kap_c ( \dtc1 + 4\dtc2 \cosh\kap_c)
        = \hf b_2, \label{a3} \\ 
b_3 &=& {\dot w}_c + 2\dtc1 \cosh\kap_c + 2\dtc2 \cosh 2\kap_c,
                   \label{b3}
\eea
and where we used \eno{Eatmc} to simplify the expression for $A_2$.
We can also simplify the results for $a_1$ and $b_3$.
Using \eno{t1c} in \eno{a1}, we obtain
\beq
a_1 = 4 t_{2c} \sinh^2 \kap_c > 0. \label{a12}
\eeq
To simplify $b_3$, we recall that the discriminant in Eq.~(\ref{cosq})
vanishes at $m=m_c$. Expanding about $m_c$, we have
\beq
t_1^2(m) - 4t_2(m) f(m) = -{16 \over J} \al^2 (m - m_c) t_{2c}^2
                             + O((m-m_c)/J)^2, \label{disc}
\eeq
where $\al$ is a positive constant.
Differentiating with respect to $m$ and setting $m=m_c$, we obtain
\beq
t_{1c}\dtc1 - 2t_{2c}{\dot f}_c - 2\dtc2 f_c =
             -{8\over J} \al^2 t_{2c}^2. \label{discdif}
\eeq
Since $f = w - 2t_2 -E$, using Eqs.~(\ref{Eatmc}) and (\ref{t1c}),
we obtain $f_c = 4 t_{2c}\cosh^2\kap_c$. Also,
${\dot f}_c = {\dot w}_c - 2\dtc2$, and the left hand side of
Eq.~(\ref{discdif}) becomes
\beq
-2 t_{2c} ({\dot w}_c + 2 \dtc1 \cosh\kap_c
            + 2\dtc2 \cosh 2\kap_c), \label{t2cb3}
\eeq
which by Eq.~(\ref{b3}) we recognize as $-2t_{2c} b_3$. Therefore,
\beq
b_3 = 4 \al^2 t_{2c}/J. \label{b32}
\eeq
One can also similarly show that
\beq
a_3 = t_{2c}\sinh\kap_c
        \left. {d \over dm}
              \left( {t_1 \over t_2} \right)\right|_{m=m_c},
          \label{a32}
\eeq
but this result is not particularly useful.

Let us now examine the order of magnitude of the various $a$ and
$b$ coefficients just introduced. We first note that
since $t_{\al c} = O(J^0)$ and ${\dot t}_{\al c} = O(J^{-1})$,
Eq.~(\ref{discdif}) implies 
that $\al = O(1)$. It follows that $a_1 = O(J^0)$, while
$a_2$, $a_3$, $b_2$, and $b_3$ are all of order $J^{-1}$.

To solve the differential equation (\ref{diffy}) with the
approximations (\ref{A0a1})--(\ref{A2a3}) for the $A$'s,
we eliminate the first derivative via the substitution
\beq
y(m) = z(m) \exp -{1\over 2a_1}
             \left[a_2(m-m_c) -
               {b_2 \over 2}\sig_2 (m-m_c)^2 \right].
                     \label{ytoz}
\eeq
Then $z(m)$ obeys
\beq
a_1 {\ddot z}(m) + b'_3 (m - m'_c) z_m = 0,
     \label{diffz}
\eeq
where we have dropped a term of order $(m - m_c)^2/J^2$ in the
coefficient of $z_m$, and where
\bea
b'_3 &=& b_3 + \sig_2 {a_2 b_2 \over 2 a_1}, \label{b3pr} \\
m'_c &=& m_c + {a_2^2 \over 4a_1 b_3}. \label{mprc}
\eea
This is Airy's differential equation, and the general solution can be
written as a linear combination of $\Ai(\zeta')$ and $\Bi(\zeta')$,
where
\beq
\zeta' = -\left( {b'_3 \over a_1}\right)^{1/3} (m - m'_c).
       \label{zetpr}
\eeq

We can now assess the validity of our approximation for $y(m)$.
First, let us ask for the order of $\dot y$ relative to $y$. From the
known behavior of the Airy functions, this is $J^{-1/3}$ in the
failure zone $ \Dta m \le J^{1/3}$, and of order $(\Dta m/J)^{1/2}$
in the overlap zone $J^\eta \gg \Dta m \gg J^{1/3}$. We thus see that
the higher order terms in Eqs.~(\ref{A0a1})--(\ref{A2a3}) are indeed
smaller than those retained. In the same way the terms dropped in the
differential equation for $z(m)$ can be seen to be small.
Second, the $d^3y/dm^3$ term which was
neglected in the differential equation (\ref{diffy}) is of order
$(\Dta m)^{3/2}/J^{1/2}$ in the overlap zone. Since $\eta < 1$, this
term is smaller than the $\ddot y$ term, and higher order derivatives
are smaller still.

The next step is to match the solution (\ref{Ctoy}) and (\ref{ytoz})
using the known asymptotic forms of the Airy functions $\Ai$ and $\Bi$,
onto the DPI forms for $m < m_c$ and $m > m_c$. The matching zones
can be taken to be any regions in which
$J^{\eta} \gg \Dta m \gg J^{1/3}$,
where $1 > \eta > 1/3$. The leading behavior of the Airy functions is
either an exponential or a sine or cosine of $(\Dta m)^{3/2}/J^{1/2}$.
We can ensure that the term $b_2(m-m_c)^2/4a_1$ in the exponential in
Eq.~(\ref{ytoz}) is inconsequential on both sides if we choose
$\Dta m \ll J^{1/2}$. Accordingly we take the matching zone as
\beq
J^{1/2} \gg |m - m_c| \gg J^{1/3}, \label{mzone}
\eeq
In this zone we can approximate $y(m) \approx z(m)$, and also
neglect the small corrections in Eqs.~(\ref{b3pr}) and (\ref{mprc}).
This amounts to saying that
\beq
C_m \approx e^{-\sig_2 \kap_c (m - m_c)} 
             [c_1 \Ai(\zeta) + c_2 \Bi(\zeta)], \label{CAiry}
\eeq
where $c_1$ and $c_2$ are two arbitrary constants, and
\beq
\zeta = -\left( {b_3 \over a_1}\right)^{1/3} (m - m_c).
       \label{zeta}
\eeq

Let us first match Eq.~(\ref{CAiry}) to the DPI solutions for
$m < m_c$. In order to treat all four solutions at the
same time, we rewrite Eq.~(\ref{cosq}) as
\beq
\cos q(m) = {-t_1(m) + \sig_1 [t^2_1(m) - 4t_2(m) f(m)]^{1/2}
              \over 4t_2(m)}, \quad (m < m_c)\label{cosq2} 
\eeq
where $f(m) = w(m) - 2t_2(m) - E$, and $\sig_1 = \pm 1$.
The wavevector $q(m)$ thus depends on both $\sig_1$ and $\sig_2$,
and hence, so does the velocity $v(m)$. We do not bother to write
down these forms explicitly, except to note that $i\sss v(m)$ is
positive. The DPI solutions may therefore be written as
\beq
C_m = {A_{\sss} \over 2\sqrt{i\sss v(m)}}
         \exp i\int_{m_c}^m q(m') dm', \label{left1} 
\eeq
where $A_{\sss}$ is a real constant which is introduced as
a notational aid in distinguishing the different cases.
In the matching zone, it follows from Eqs.~(\ref{cosq2},
(\ref{defkc}), (\ref{disc}), and (\ref{vm}), that
\bea
\cos q(m) &\approx& \cosh \kap_c 
                       +  \al\,\sig_1 ((m_c - m)/J)^{1/2}, \nnu \\
q(m) &\approx& i\sig_2\kap_c
          + i \sss {\al \over \sinh\kap_c}((m_c - m)/J)^{1/2}
            \label{mltmc}, \\
v(m) &\approx& -8i\al\sss \sinh\kap_c\,|t_{2c}|\,
                       ((m_c - m)/J)^{1/2}. \nnu
\eea
(We have written $|t_{2c}|$ instead of $t_{2c}$ with a view to
including the case $t_1 < 0$, $t_2 < 0$, which we shall consider
later.) Therefore,
\beq
C_m    \approx {A_{\sss} \over 4\sqrt{2\al t_{2c} \skc}}
       \left({m_c -m \over J}\right)^{-1/4}
       \exp\left[ -\sig_2\kap_c (m - m_c)
                +{2 \over 3}{\al\sss \over\skc}
                 {(m_c - m)^{3/2} \over J^{1/2}} \right].
                 \label{left2}
\eeq
Since $\zeta \gg 1$ for $m < m_c$ in the zone (\ref{mzone}), we may
use the asymptotic forms \cite{as}
\bea
\Ai(\zeta) &\approx& {1 \over 2\sqrt\pi} \zeta^{-1/4}\
            \exp \lp -{2\over 3}\zeta^{3/2} \rp, \label{Aileft} \\
\Bi(\zeta) &\approx& {1 \over \sqrt\pi} \zeta^{-1/4}\
            \exp \lp {2\over 3}\zeta^{3/2} \rp. \label{Bileft}
\eea
Using Eq.~(\ref{zeta}) and comparing with Eq.~(\ref{left2}), we
see that we must use the function $\Ai$ when $\sig_2 = -\sig_1$,
and $\Bi$ when $\sig_2 = \sig_1$. We can write
\beq
C_m = KA_{\sss} e^{-\sig_2\kap_c(m-m_c)}\times\cases
             {\Bi(\zeta), & $\sig_2 = \sig_1$; \cr
              2\Ai(\zeta), & $\sig_2 = -\sig_1$, \cr}
               \label{CAiBi}
\eeq
with
\beq
K = {\sqrt\pi  \over 4\sqrt{2\al t_{2c}\skc}}
             \lp{a_1 \over b_3} \rp^{-1/12} J^{1/4}
       \label{Kdef}
\eeq

To connect Eq.~(\ref{CAiBi}) with the DPI solutions for $m > m_c$,
we make use of the reality principle, namely that the solution
to the basic recursion relation (\ref{Seq}) must be real for
$m>m_c$ if it is real for $m < m_c$. Since the solutions to
Eq.~(\ref{cosq}) now lead to complex $q$, it is clear that we must
add two DPI solutions in order to get a real result. To this end
we define
\beq
q_a(m) = i\sig_2 \kap(m) + \chi(m), \quad m > m_c
   \label{qmrite}
\eeq
where $\kap(m)$ and $\chi(m)$ are both real and positive,
$\kap(m_c) = \kap_c$, and
\bea
\cosh\kap \cos\chi &=& -t_1/4|t_2|, \label{chcos} \\
\sinh\kap \sin\chi &=& (4t_2f - t_1^2)^{1/2}/4|t_2|. \label{shsin}
\eea
In terms of these quantities, we have
\bea
\cos q_a(m) &=& \cosh\kap(m)\cos\chi(m) -
    i\sig_2\sinh\kap(m) \sin\chi(m), \label{cosqr} \\
s_a(m) &\equiv& -i\sig_2 v_a(m) =  8 |t_2(m)| \sinh\kap(m)
                   \sin\chi(m) \sin q_a(m). \label{sam}
\eea
[As before, we have written $|t_2(m)|$ instead of $t_2(m)$
with a view to subsequently including the case $t_1 < 0$, $t_2 < 0$.] 
The corresponding quantities for the complex conjugate solution
are given by $q_b(m) = q_a^*(m)$, $s_b(m) = s_a^*(m)$. The DPI
solution into which Eq.~(\ref{left1}) continues for $m > m_c$
may therefore be written as
\beq
C_m = \hf B_{\sss}
             \left[ {1 \over \sqrt{s_a(m)}}
         \exp \left(i\int_{m_c}^m q_a(m') dm'
           + i \Dta_{\sss} \right) + {\rm c.c.} \right]
                    \label{right1}
\eeq
where $B_{\sss}$ is real, $\Dta_{\sss}$ is a
phase to be found, and c.c. denotes ``complex conjugate".

It remains to compare the asymptotic forms of Eqs.~(\ref{CAiBi})
and (\ref{right1}) in the overlap zone (\ref{mzone}), and thus
find $\Dta_{\sss}$ and the relation between $A_{\sss}$
and $B_{\sss}$. Using
the asymptotic forms for $\Ai$ and $\Bi$, we have
\beq
C_m \approx K A_{\sss}
           {e^{-\sig_2\kap_c(m-m_c)} \over \sqrt\pi}
            (-\zeta)^{-1/4} \times \cases
            {\cos\left[ {2\over 3}(-\zeta)^{3/2} + {\pi\over 4}\right],
                 & $\sig_2 = \sig_1$; \cr
            2\sin\left[ {2\over 3}(-\zeta)^{3/2} + {\pi\over 4}\right],
                 & $\sig_2 = -\sig_1$. \cr}
                  \label{Cright}
\eeq
To write Eq.~(\ref{right1}) in a similar form, we first note that
in the matching zone,
\bea
q_a(m) &\approx& i\sig_2 \kap_c +
         \sig_2 {\al \over\skc} ((m-m_c)/J)^{1/2}, \label{qam} \\
s_a(m) &\approx& 8 t_{2c} \al \skc \lp {m - m_c \over J} \rp^{1/2}
             \exp\left[ i\sig_2
                 {\pi\over 2} - i\sig_2\al {\ckc \over \sinh^2\kap_c}
                          \lp {m - m_c \over J} \rp^{1/2} \right].
                  \label{sam2}
\eea
Therefore,
\beq
i\int_{m_c}^m q_a(m') dm' = -\sig_2\kap_c \Dta m
                            + i {2 \over 3} (-\zeta)^{3/2},
                          \label{intqa}
\eeq
where we have used Eqs.~(\ref{a12}), (\ref{b32}), and (\ref{zeta})
to simplify the term in $(-\zeta)^{3/2}$.
Compared to this term,
the correction proportional to $(\Dta m)^{1/2}$ in the exponent for
$s_a(m)$ is of higher order in $1/J$ in the matching zone, and can be
ignored. Doing this, and making use of Eqs.~(\ref{zeta}) and
(\ref{Kdef}), we get
\beq
C_m \approx {K \over \sqrt\pi} B_{\sss} (- \zeta)^{-1/4}
       e^{-\sig_2\kap_c \Dta m}
             \left[ \exp\lp i {2\over 3}(-\zeta)^{3/2}
                 - i{\pi\sig_2\over 4} + i \Dta_{\sss} \rp
                  + {\rm c.c.} \right].
    \label{right2}
\eeq 
Comparing with Eq.~(\ref{Cright}), we see that
$B_{\sig_1\sig_1} = A_{\sig_1\sig_1}/2$, while
$B_{\sig_1,-\sig_1} = A_{\sig_1,-\sig_1}$. Further,
$\Dta_{\sss}$ vanishes if $\sig_1 = -1$, and equals $\pm\pi/2$
if $\pm\sig_2 = \sig_1 = 1$. These results can be summarized as
\bea
B_{\sss} &=& {2 - \delta_{\sss} \over 2}
                     A_{\sss},
                       \label{BtoA} \\
\Dta_{\sss} &=& {\pi \over 4}(1+\sig_1)\sig_2.
                        \label{Maslov}
\eea

Putting together Eqs.~(\ref{left1}), (\ref{right1}),
(\ref{BtoA}), and
(\ref{Maslov}), the connection formula may be written in the final
form
\bea
&&{A_{\sss} \over 2\sqrt{i\sss v(m)}}
        \exp i\int_{m_c}^m q(m') dm' 
                     \gets C_m \nnu \\
                 &\qquad&\to
      \lp1 - \hf\delta_{\sss}\rp A_{\sss} 
             \left[ {1 \over \sqrt{s_a(m)}}
               \exp \left(i\int_{m_c}^m q_a(m') dm'
                 + i {\pi\over 4}(1+\sig_1)\sig_2 \right)
                   + {\rm c.c.} \right].
    \label{conform}
\eea
The wavevectors $q$ and $q_a$, and the velocity and speed,
$v(m)$ and $s_a$, to be used
for a given set of signs $(\sig_1, \sig_2)$ are given by
Eqs.~(\ref{mltmc}) and (\ref{qam}).

We use the double arrow notation $\gets C_m \to$ advocated by
Heading \cite{jh} and Dingle \cite{rbd}
to emphasize the bidirectionality of the connection formula.
We refer readers
to these authors for lucid discussions of this point, but since even
as masterly and authoratative a text as Landau and Lifshitz \cite{ll}
states
that connection formulas may only be used in one direction, it may be
worth paraphrasing their remarks briefly. Thus, as
stated by Heading, the notation
means that there {\it is} a solution $C_m$ to the recursion relation
(\ref{Seq}) with the stated asymptotic behaviors for $m < m_c$
and $m > m_c$. And, as stressed by Dingle, the formula merely states
that a certain exponentially growing on one side of the turning point,
{\it which is free from the growing component}, goes over into a
certain oscillatory solution (with an exponentially growing envelope)
on the other side, and vice versa. Likewise for the growing component.
It says nothing about whether or not we can use the formula in both
directions in physical problems where we do not have complete
information. If, e.g., a solution with $\sig_1 = \sig_2 = -1$
for $m < m_c$ contains an admixture of the $\sig_1 = -\sig_2 = 1$
solution of small but indeterminable magnitude, then we can
say nothing about the amplitude or the phase of the oscillatory
factor for $m > m_c$. It is in these situations, where information
is incomplete, that the reservations about the unidirectionality
of the connection formulas are relevant.

The connection formula for the case where $t_1 < 0$ and $t_2 < 0$
can be derived in exact parallel by making minor modifications to the
intermediate steps. The final form is sufficiently different to be
worth giving separately:
\bea
&&{A_{\sss} e^{i \pi m_c} \over 2\sqrt{i\sss v(m)}}
         \exp i\int_{m_c}^m q(m') dm' 
 \gets C_m \nnu \\
         &\qquad& \to
      \lp1 - \hf\delta_{\sss}\rp A_{\sss}
             \left[ {e^{i\pi m_c} \over \sqrt{s_a(m)}}
               \exp \left( i\int_{m_c}^m q_a(m') dm'
                 - i {\pi\over 4}(1-\sig_1)\sig_2 \right)
                   + {\rm c.c.} \right].
    \label{conform2}
\eea
The quantities $v(m)$ and $s_a(m)$ are given by the same formal
expressions as before, but now
\bea
q(m) &\approx& \pi + i\sig_2\kap_c
          + i \sss {\al \over \sinh\kap_c}((m_c - m)/J)^{1/2},
            \label{qm2} \\
q_a(m) &=& \pi + i\sig_2 \kap(m) + \chi(m). \label{qam2}
\eea
In Eq.~(\ref{qam2}), $\kap(m)$ and $\chi(m)$ are also given by
the same formal expressions as before, i.e., Eqs.~(\ref{chcos})
and (\ref{shsin}).

\acknowledgments
This work is supported by the NSF via grant number DMR-9616749.
I am indebted to Wolfgang Wernsdorfer and Jacques Villain for
useful discussions and correspondence about \Fe8.

\appendix
\section{Width of failure zone of DPI approximation}
\label{valid}

We have seen in \Slab{basic} that the DPI approximation is valid
everywhere except near turning points. Let us now estimate the size
of the region where it fails.
Suppose that $q=0$ or $\pi$ at the turning point, $m_c$. As shown in
Eqs.~(\ref{qntp}) and (\ref{vntp}), $q(m)$ and $v(m)$ both vary as
$[(m-m_c)/J]^{1/2}$ for $m$ near $m_c$. Thus
$\ddot q \sim (m-m_c)^{-3/2} J^{-1/2}$, and the first term in $\Phi_2$
is of order $[(m-m_c)J]^{-1/2}$. Next, note that
$r \sim [J/(m-m_c)]^{1/2}$, so that the second and third terms are both
of order $J^{1/2}(m-m_c)^{-3/2}$. The first term is subdominant to these
two, and therefore, the condition that $\Phi_2 \ll 1$ reduces to
\beq
|m - m_c| \gg J^{1/3}. \label{mgg}
\eeq

In the case where the second factor in $v(m)$ vanishes, we have
$t_1 + 4t_2 \cos q \sim [(m-m_c)/J]^{1/2}$, and
$q - q_c  \sim [(m-m_c)/J]^{1/2}$. Thus the integrand of the
first term in $\Phi_2$ varies as $1/(m-m_c)^2$, and the term itself
is of order $1/(m-m_c)$. However,  
$r \sim [J/(m-m_c)]^{1/2}$ as before, so that the second and third terms
are again of order $J^{1/2}(m-m_c)^{-3/2}$, and much greater than the
first term. The condition for validity of the DPI form is again given
by Eq.~(\ref{mgg}).

The condition can be written in alternative form by noting that near the
region of failure, $r(m) \sim 1/v(m)$, and that $\dot v \sim \dot q$.
Thus $dr/dm \sim \dot q/v^2$, and the last term in $\Phi_2$ can
also be shown to be of order $\dot q/v^2$. Thus the condition for
validity can be written as
\beq
\dot q(m) \ll v^2(m). \label{qdllv2}
\eeq

\begin{figure}
\caption{Schematic band energy curves arising from five term recursion
relations. Curve 1 arises if $|t_1/4t_2| > 1$, curve 2 if
$-1< t_1/4t_ <0$, and curve 3 if $0 < t_1/4t_2 < 1$.  The dual
labelling scheme for the critical energies $U_0$, $U_\pi$, etc. is
shown.}
\label{bands}
\end{figure}

\begin{figure}
\caption{Critical energy curves in the case where
$-t_1(m)/4t_2(m) < -1$ for all $m$. The points $m=m_b$ and $m=m_c$
are turning points of type $\bar A$ and $A$, respectively, for the
energy $E$.}
\label{ccurves}
\end{figure}

\begin{figure}
\caption{Same as figure 2, but with 
$-t_1(m)/4t_2(m) \in [-1,1]$ for some $m$. These critical curves arise
from the Hamiltonian (\ref{ham}) for \Fe8. Note that $U_*$ is the
lower band edge $U_-$ for the central $m$ region, and the forbidden
critical energy $U_f$ for the outer $m$ regions. Correspondingly,
$U_0$ is the internal critical energy $U_i$ in the central region,
and the lower band edge $U_-$ in the outer regions. The point $m=m_c$
is a type $A'$ turning point for the energy $E$.}
\label{critfe8}
\end{figure}

\begin{figure}
\caption{Magnified view of lower left hand portion of \fno{critfe8}.
The points $m= m_t$ and $m_c$ are turning points of type $A$ and
$B$ respectively.}
\label{fe8critmag}
\end{figure}

\begin{figure}
\caption{Band energy curves in the vicinity of a type $A'$ turning
point.}
\label{Aprband}
\end{figure}

\begin{figure}
\caption{Possible band energy curve for a recursion relation with
seven or more terms.}
\label{strband}
\end{figure}

\begin{figure}
\caption{Critical curves for a system with band energy curves
as shown in \fno{strband}.}
\label{strcrit}
\end{figure}

\begin{references}
\bibitem[*]{byline} Electronic address: agarg@nwu.edu

\bibitem{dm} R.~B. Dingle and J. Morgan, Appl.\ Sci.\ Res.\ {\bf 18},
221 (1967); {\it ibid.} {\bf 18}, 238 (1967).

\bibitem{sg} K. Schulten and R.~G. Gordon,
J.\ Math.\ Phys.\ {\bf 16}, 1971 (1975).

\bibitem{pab} P.~A. Braun, Rev.\ Mod.\ Phys.\ {\bf65}, 115 (1993).

\bibitem{pab2} P.~A. Braun, Teor.\ Mat.\ Fizika {\bf 37}, 355
(1978) [Sov.\ Phys.\ Theor.\ Math.\ Phys.\ {\bf 37}, 1070 (1978)].

\bibitem{vs} J.~L. van Hemmen and A.~S\"ut\H o,
(a) Europhys.\ Lett.\ {\bf 1}, 481 (1986);
(b) Physica {\bf 141B}, 37 (1986).

\bibitem{cs} C.~Sangregorio, T.~Ohm, C.~Paulsen, R.~Sessoli,
and D.~Gatteschi, Phys. Rev. Lett. {\bf 78}, 4645 (1997).

\bibitem{alb}
A.-L. Barra, P.~Debrunner, D.~Gatteschi, Ch.~E. Schultz, and R.~Sessoli,
Europhys.\ Lett.\ {\bf 35}, 133 (1996).

\bibitem{rc} R.~Caciuffo, G.~Amoretti, A.~Murani, R.~Sessoli,
A.~Caneschi, and D.~Gatteschi, Phys.\ Rev.\ Lett.\ {\bf 81}, 4744 (1998).

\bibitem{bw} M.~V. Berry and M.~Wilkinson, Proc.\ Roy.\ Soc.\ Lond.\ A
{\bf 392}, 15 (1984).

\bibitem{agepl} A.~Garg, Europhys.\ Lett.\ {\bf 22}, 205 (1993).

\bibitem{agdia} A.~Garg, submitted to Europhysics Letters.

\bibitem{vf} J.~Villain and Anna Fort, submitted to Europhys. B.

\bibitem{ws} W.~Wernsdorfer and R.~Sessoli, Science {\bf 284}, 133 (1999).

\bibitem{agprb} A.~Garg, Phys.\ Rev.\ B {\bf51}, 15161 (1995).

\bibitem{prl99} A.~Garg, Phys.\ Rev.\ Lett.\ {\bf 83}, 4385 (1999);
and submitted to Phys.\ Rev.\ B.

\bibitem{fn1} There are clearly situations where it is better for
$m$ to take on non-integer values, or to change in increments other
than unity. For developing the formalism, however, little is gained by
incorporating these refinements, and it is sufficient to
let $m$ take on contiguous integer values.

\bibitem{fn3} Note that it is possible for more than one {\it pair} of
$q$'s to coalesce at a turning point, as in all cases with the letter
symbol $B$. For non-Hermitean problems, turning points are still
characterized by the coalescence of two roots for $q$, but the
directions in which they approach or separate can be arbitrary.

\bibitem{bm} M.~V. Berry and K.~E. Mount, Rep.\ Prog.\ Phys.\ {\bf 35},
315 (1972).

\bibitem{agjmp} A.~Garg, J.\ Math.\ Phys.\ {\bf 39}, 5166 (1998).

\bibitem{Bohm} David Bohm, {\it Quantum Theory} (Prentice-Hall,
Englewood Cliffs, 1951), Chap. 12.

\bibitem{as} {\it Handbook of Mathematical Functions},
M.~Abramowitz and I.~A. Stegun, eds. (Dover, New York, 1970),
Sec.~10.4.
 
\bibitem{jh} J.~Heading, {\it An Introduction to Phase-Integral
Methods} (Methuen, London, 1962). See Chap. 1, Sec.~1.6 in
particular.

\bibitem{rbd} R.~B. Dingle, {\it Asymptotic Expansions: Their
Derivation and Interpretation} (Academic Press, London, 1973).
See Chap. XIII, Secs.~13.3 and 13.4 in particular.

\bibitem{ll} L.~D. Landau and E.~M. Lifshitz, {\it Quantum Mechanics}
(Pergamon, New York, 1977), 3rd ed.. See Sec.~47.

\end{references}
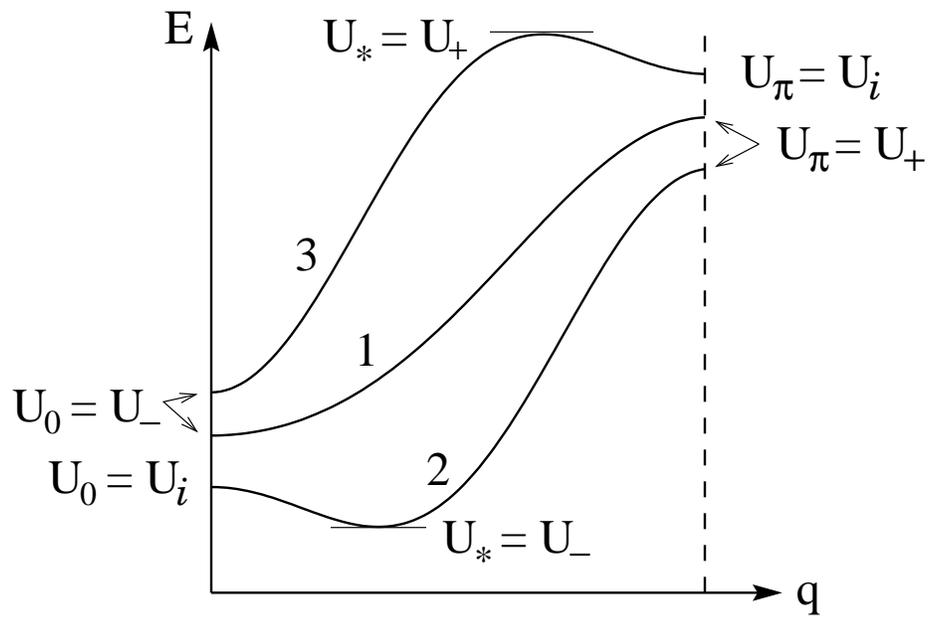
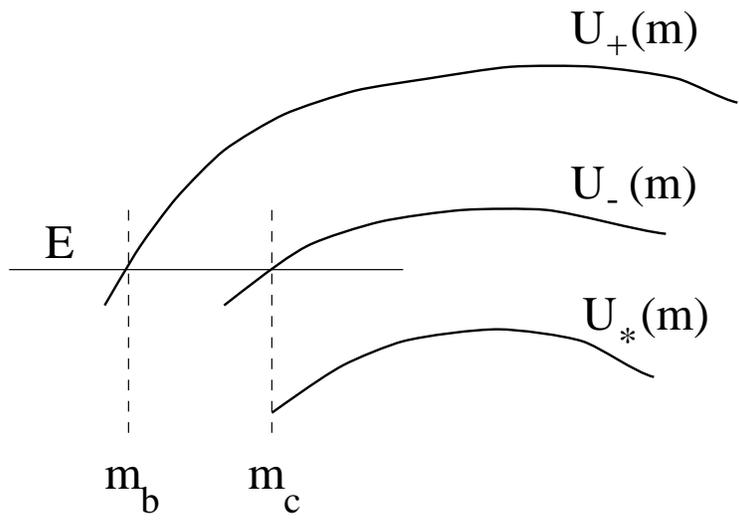
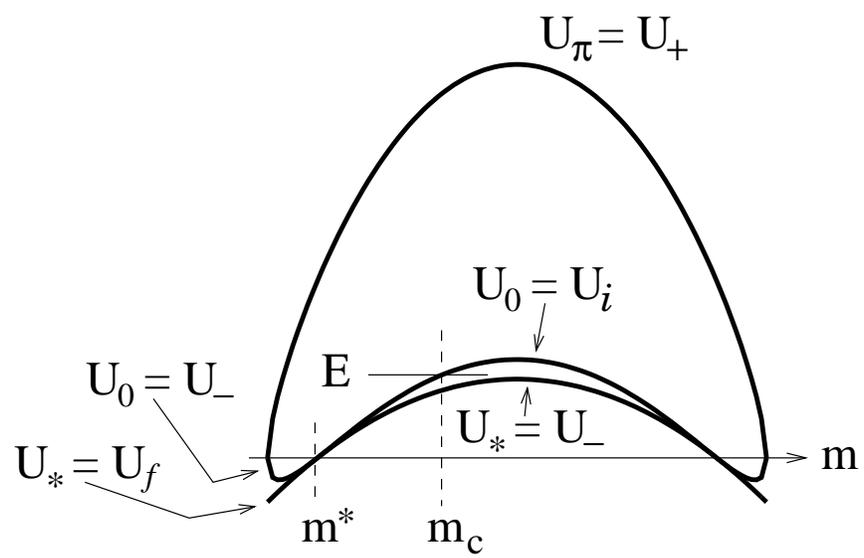
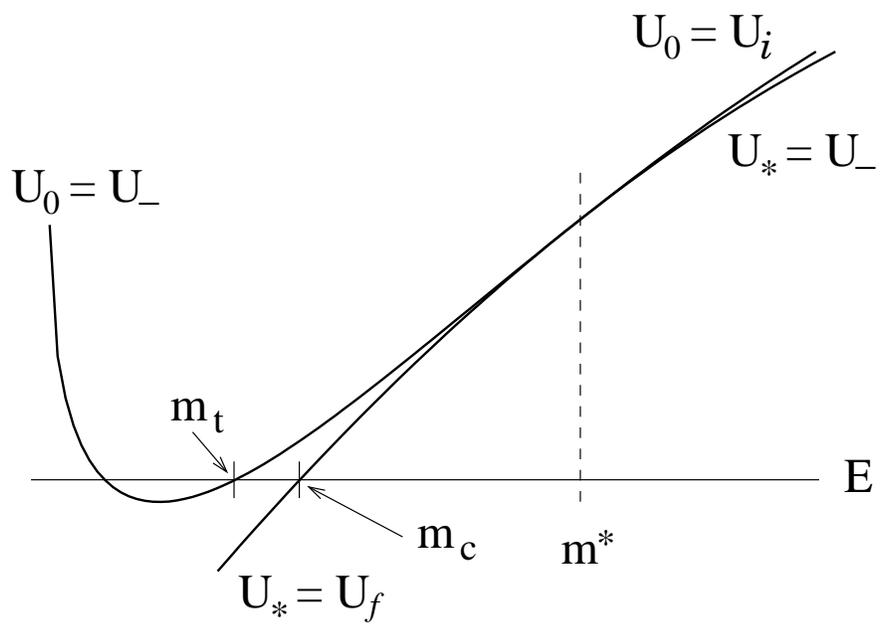
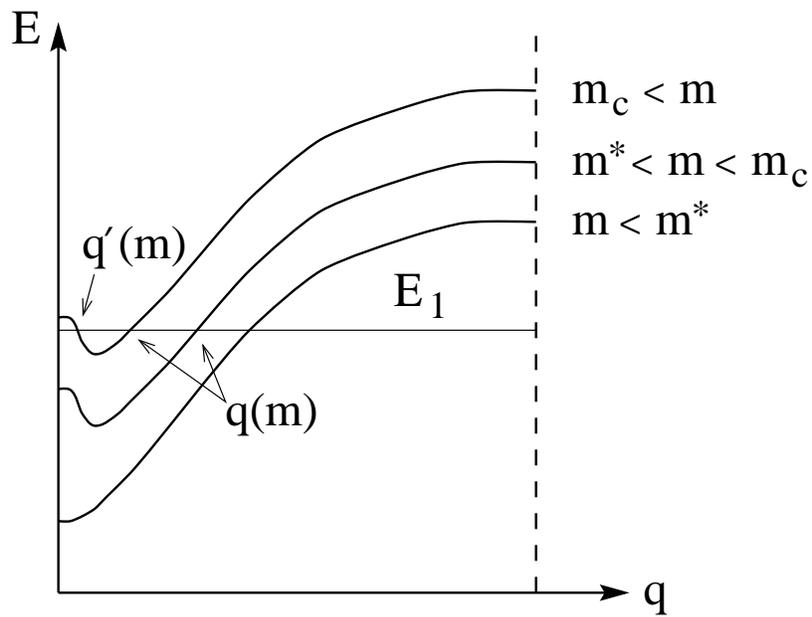
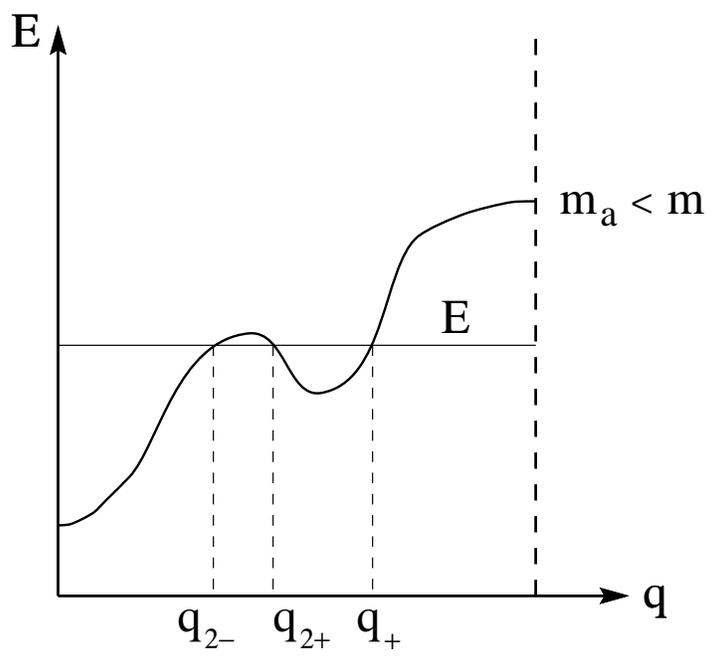
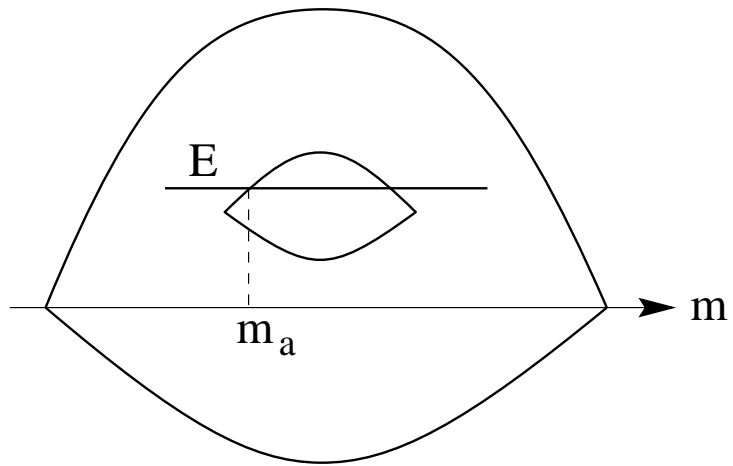
\end{document}